# Calculation of a temperature-volume phase diagram of water to inform the study of isochoric freezing down to cryogenic temperatures

Matthew J. Powell-Palm

*Department of Mechanical Engineering, Texas A&M University, College Station, TX 7783, USA*
*Department of Mechanical Engineering, University of California at Berkeley, Berkeley, CA 94720, USA*

To whom correspondence should be addressed:
MJPP : mpowellp@berkeley.edu

**Phase diagrams are integral to the application and interpretation of materials thermodynamics, and none is more ubiquitous than the common Temperature-Pressure diagram of water and its many icy phases. Inspired by recent advances in isochoric thermodynamics, we here employ a simple convex hull approach to efficiently calculate an updated Temperature-Volume phase diagram for water and five of its solid polymorphs from existing Helmholtz free energy data. We proceed to highlight fundamental similarities between this T-V diagram and conventional binary temperature-concentration (T-x) diagrams, provide the volume coordinates of a variety of three-phase invariant reactions (e.g. "confined" or "volumetric" eutectics, peritectics, etc.) that occur amongst the phases of pure water under isochoric or confined conditions, and calculate the phase fraction evolution of Ice Ih with temperature along multiple isochores of interest to experimental isochoric freezing. This work provides a requisite baseline upon which to extend the study of isochoric freezing to cryogenic temperatures, with potential applications in thermodynamic metrology, cryovolcanism, and cryopreservation.**

## Introduction

Isochoric (constant-volume or confined) freezing has recently emerged as an alternate thermodynamic modality by which to explore fundamental aqueous thermodynamics[1], ice nucleation kinetics[2], and cryopreservation[3,4]. During isochoric freezing, an aqueous system is confined within a rigid, high-strength container in the absence of bulk air or other heterogeneous phases, fixing both the absolute and specific volume of the water within and denying the system access to the pressure reservoir provided by the atmosphere[5,6]. By controlling the volume of the system rather than the pressure, the thermodynamic path the system takes as it is cools is altered fundamentally, with the system occupying various two-phase equilibrium states instead of the single-phase equilibrium states encountered when cooling under constant pressure. This process has been well characterized for the temperature range 0 to -22°C[5,6], wherein an aqueous isochoric system will exist in stable two-phase equilibrium between ice Ih and liquid water. However, the process of isochoric freezing down to lower temperatures, which will involve multiphase equilibria comprised of multiple forms of ice, has not been rigorously examined in the modern era. Here, we use a purely-geometric convex hull technique to calculate a new Temperature-Volume phase diagram for water and five of its icy polymorphs, providing an essential reference tool for future lower-temperature work in isochoric freezing. We use this diagram to highlight fundamental similarities in the phase change processes of single-component confined systems and binary unconfined systems, highlight pure-substance three-phase invariant reactions that can occur in confined aqueous systems, and calculate for the first time the phase fraction evolution of ice 1h during cooling to 190°C along multiple isochores. In sum, this work seeks to provide the baseline thermodynamic information needed to inform future studies of isochoric freezing at cryogenic temperatures.

## Geometric implications of extensive variables in phase diagrams

Phase diagrams, which map material equilibrium, may be constructed to reflect a wide variety of natural thermodynamic variables[5,7–11]. Amongst the most common diagrams for pure substances and binary substances respectively are the Temperature-Pressure and Temperature-concentration diagrams, which have been deployed widely since the turn of the 20th century. These two diagrams exhibit markedly different geometry, and are characterized by markedly different features.

The Temperature-Pressure diagram for a single-component substance is constructed by projecting the Gibbs free energy surfaces $G(T,P)$ of each phase onto $T$-$P$ axes. Because temperature and pressure are both intensive variables and thus concave in the $G$-$T$-$P$ energy space, $T$-$P$ phase diagrams produce 2-dimensional single-phase regions, bounded by 1-dimensional boundary lines (such as the liquidus curve) along which two phases may coexist and producing 0-dimensoinal three-phase coexistence points at the meetings of any three single-phase regions (so-called triple points)[11].

The Temperature-Concentration diagram for a binary substance is constructed for a given pressure (typically atmospheric) using the Gibbs free energy surfaces $G(T,x)$ of each phase. Critically however, concentration $x$ is an extensive variable (the conjugate of the intensive chemical potential $\mu$), and thus introduces convexity in the $G$-$T$-$x$ energy space, as guaranteed by the fundamental thermodynamic stability criterion $\frac{\partial^2 U}{\partial X^2} \geq 0$, in which $U$ is the internal energy of the substance and $X$ is an arbitrary extensive variable[12].

As Gibbs demonstrated in his seminal *On the Equilibrium of Heterogeneous Substances*[13], this convexity introduces the opportunity for a system to reduce its free energy by forming a heterogeneous mixture of two or more co-existing phases across a range of values along the convex axis (here concentration), defined by the tangent line in $G$-$x$ space that is common to each of the phases. Following the advent of modern convex geometry theory in the 1960s, it has become standard to map the myriad common tangents that join single-phase regions and define multi-phase regions by taking the convex hull of the free energy surfaces of each phase[5,14,15]. It is the projection of the lower envelope of this convex hull (as opposed to the simple projection of the energy surfaces themselves) onto axes of temperature and concentration that then provides the standard binary $T$-$x$ phase diagram.

As a consequence of this convexity of the free energy surfaces, binary $T$-$x$ diagrams possess many features that the $T$-$P$ diagrams of pure substances do not. Most evidently, they possess both 2D single-phase regions *and* 2D two-phase coexistence regions, alongside 1D three-phase coexistence lines as opposed to the 0D three-phase triple points found in $T$-$P$ diagrams. These varied heterogeneous equilibrium regions are also accompanied by various reactions or transitions that describe the unique passage between them. Most notable of these reactions is the eutectic transition, which describes when a single liquid phase transitions into a two-phase solid-solid mixture and defines the lowest temperature at which the liquid phase is stable. This transition or reaction is often described as a "three-phase invariant", as the transition to the solid-solid mixture requires traversing the three-phase, liquid-solid-solid coexistence line (i.e. the eutectic line). Other three-phase invariant reactions include the peritectic, monotectic, eutectoid, peritectoid, etc. These reactions are integral features of multicomponent phase diagrams, but are seldom recognized in pure systems (as reinforced by the absence of such features on the pure-substance $T$-$P$ diagram).

## Calculation of a Temperature-Volume phase diagram for six condensed phases of water

Here, we demonstrate that these same three-phase reactions may and will also occur in pure-substances under isochoric or confined conditions. We begin by constructing the Temperature-Volume phase diagram for water in the temperature range 190 – 353.5 K, including the single-phase equilibrium regions of liquid water, ice-1h, ice-II, ice-III, ice-V, and ice-VI. This temperature range was chosen based on the phases for which mutually-consistent equations of state were available. To construct this phase diagram, we extend the basic method of Powell-Palm, Rubinsky & Sun[5]: the 2D convex hulls enclosing the Helmholtz free energy surfaces $F(T,V)$ of all of the aforementioned phases were calculated at discrete temperatures within the stated range (Fig. 1a), with an increment of 0.0654 K; single-phase points on the lower convex hull (Fig. 1b), which represent equilibrium states, were projected onto axes of the natural variables of the system $(T,V)$; and two-phase coexistence regions and three-phase coexistence lines, which definitionally join the single-phase regions within the convex hull, were identified and marked accordingly. See the Methods section for further information on both the algorithms used to calculate the stable single-phase regions and to identify the three-phase invariant lines. All material data were generated via the SeaFreeze framework[16].

The final $T$-$V$ diagram is presented in Figure 1.c. To the author's knowledge, this diagram represents the first calculation of the isochoric phase equilibria of water since the empirical P-V-T diagram of Verwiebe in 1939[17], and should provide a rigorous reference for future studies of isochoric freezing.

It should be noted further that all calculations have been performed on a per-kilogram basis, i.e. the $x$-axis of the diagram features specific as opposed to absolute volume. The author will emphasize that, as defined by Gibbs, Callen, and many other seminal materials thermodynamicists, the volume remains an *extensive* thermodynamic variable even in its specific form, as the product of specific volume and its conjugate intensive variable, pressure, still yields units of (specific) energy (J/kg). The ratio of any extensive variable (e.g. volume) to any other extensive variable (e.g. mass) does not alter the extensive nature of the variable in a thermodynamic sense, but instead merely provides a convenient density of extensive values. This principal is evident most commonly in the concentration, which is a density of the mass of one component against the mass of another, and provides the extensive conjugate to the intensive chemical potential. Henceforth in this work, for the purposes of thermodynamic analysis, the volume, whether absolute or specific, will be referred to as extensive.

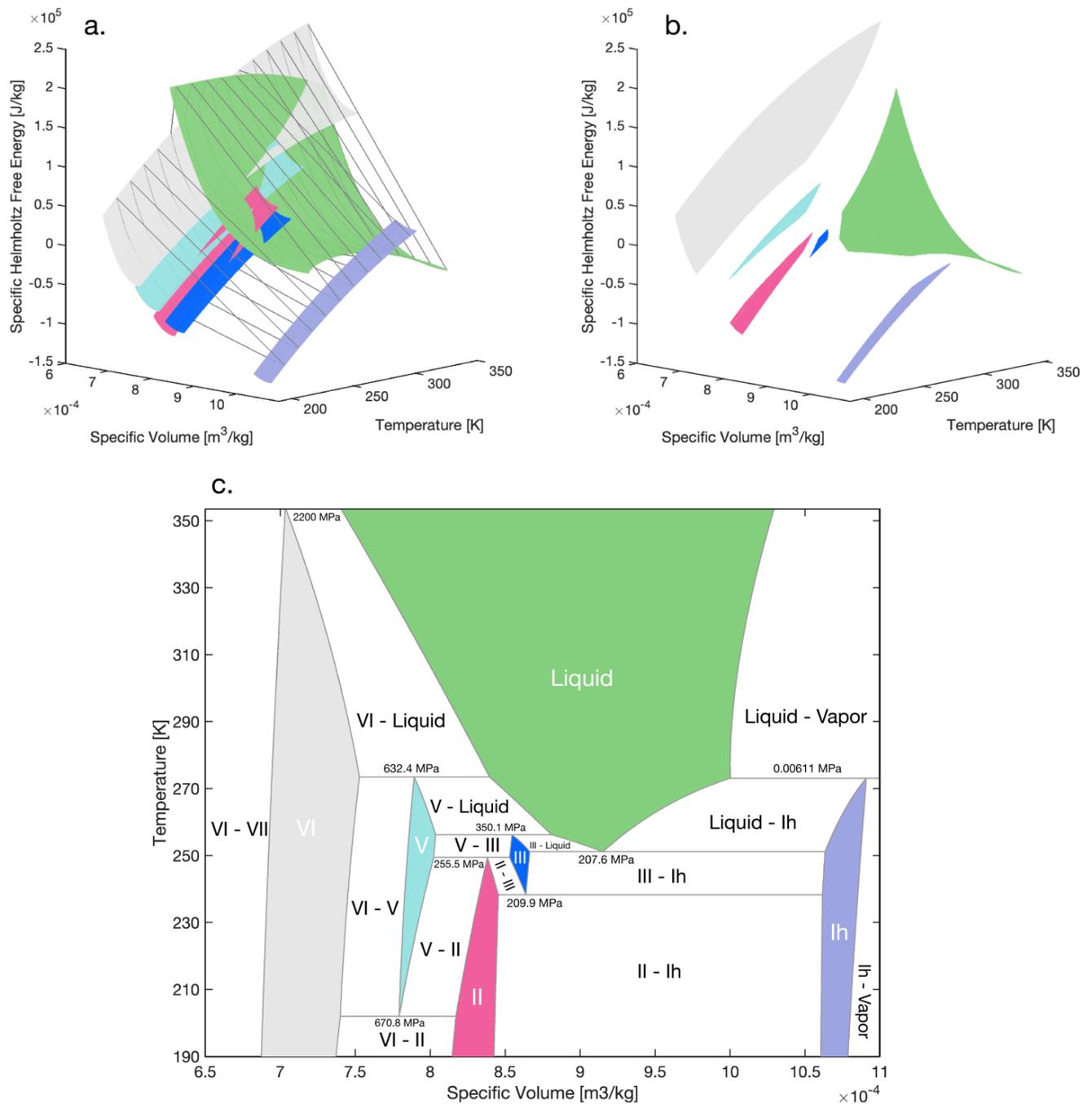

**Figure 1. Calculation of a new Temperature-Volume diagram for water. a)** This diagram was constructed by first plotting the 3-dimensional Helmholtz free energy surfaces $F(T,V)$ of each phase in temperature-volume space, then constructing a series of 2-dimensional convex hulls around these phases in axes of free energy – volume $F(V)$ as shown, evaluated at isotherms spaced by 0.08K. **b)** The single-phase points that rest on the bottom portion of the convex hull, which define the stable single phase regions, are then isolated from all other points, which represent metastable or unstable states at a given temperature-volume coordinate. **c)** Finally, the lower convex hull points shown in (b) are projected onto axes of temperature and volume, producing the T-V phase diagram. All free energy data were generated using the SeaFreeze framework. Two-phase co-existence regions are labeled, and horizontal lines mark three-phase coexistence lines, the corresponding pressures of which are also labeled

**Three-phase invariant reactions in pure systems under isochoric conditions**

Important morphological aspects of the *T-V* diagram of water are noted as follows: Firstly, the reader will notice that this *T-V* diagram appears identical in structure to conventional binary *T-x* diagrams, even though this diagram describes a pure substance. This similarity is a consequence of the fundamental similarity between all extensive and intensive variables in Gibbsian thermodynamic geometry—free energy surfaces are convex in extensive variables, leading to heterogeneous or multiphase equilibrium, and concave in intensive variables, leading to homogeneous or single-phase equilibrium[13,18]. Thus, enforcing volume (extensive) as a natural variable of the system is directly thermodynamically analogous to enforcing concentration (extensive) as a natural variable of the system. In this way, we can understand by analogy not only the two-phase equilibrium regions, the phase composition of which may be analyzed by the Lever Rule as in a binary phase diagram, but also the three-phase invariant lines. These lines, which occur at the temperatures *and pressures* of the triple points common in *T-P* phase diagrams, are directly analogous to three-phase invariant lines in binary diagrams; they mark a line in the extensive axis (here volume) along which one may move at a given temperature and maintain three-phase coexistence. Furthermore, the point at which a single-phase region (such as the liquid water region) touches the three-phase coexistence line provides a direct analogue to the binary eutectic point. At the volume coordinate of this point, one may cool directly from liquid water to an ice Ih – ice-III two-phase mixture for example, bypassing the liquid-ice Ih and liquid-ice III two-phase coexistence regions. To the author's knowledge, the premise of eutectic-like transitions in pure substances under isochoric conditions remains unexplored in contemporary literature on isochoric freezing, and we will thus take the liberty to label the aforementioned eutectic-like point as the "volumetric eutectic". The volumetric eutectic point of liquid water as calculated herein occurs at 9.16 x $10^{-4}$ $m^3$/kg and 251.2 K.

Additionally, as stands to reason, other features analogous to binary phase diagrams can be found in Figure 1.c, including volumetric eutectoid points (e.g. from ice III to the ice II – ice Ih two-phase mixture [8.64×$10^{-4}$ $m^3$/kg, 238.2 K]); volumetric peritectic points (e.g. from the ice VI – liquid water two-phase mixture to single-phase ice V [7.89×$10^{-4}$ $m^3$/kg, 273.3K]); and volumetric peritectoid points (e.g. from the ice V – ice III two-phase mixture to single-phase ice II [8.38×$10^{-4}$ $m^3$/kg, 249.4 K]).

**Isochoric phase fraction evolution when traversing multiple multiphase regions**

One integral thermodynamic property of isochoric systems (and indeed of all systems with an enforced extensive thermodynamic variable) is the phase fraction, which describes the equilibrium phase composition of a multiphase system at constant volume. Unlike other aspects of isochoric phase equilibria (such as the melting point or the three-phase invariant temperature), the phase fraction is *not* inferable from the Temperature-Pressure phase diagram, as the enforcement of intensive variables rather than extensive removes all forced phase heterogeneity from the system[13].

The evolution of the ice Ih phase fraction with temperature (and pressure) has been calculated previously for the ice 1h - liquid water two-phase equilibrium region between 0°C and approximately -22°C, using both iterative mechanical-equilibrium techniques[6] and convex hull-based techniques[5]. To the author's knowledge, the phase fractions at play in other multiphase regions have not been previously calculated, and we thus proceed to calculate them here.

Using the *T-V* phase diagram, the volumetric phase fraction *f* along a given isochore can be calculated directly using the Lever Rule:

$$f = \frac{(v_s - v_l)}{(v_r - v_l)}$$

wherein, at a chosen volume coordinate $v_s$ and at a given temperature, $v_r$ is the minimum volume of the less-dense phase to the right of $v_s$ and $v_l$ is the maximum volume coordinate of the denser phase to the left of $v_s$.

From a practical experimental perspective, $v_s$ is the system volume at which the initial mass of water (nigh-exclusively in the liquid state) is confined. Most experimental isochoric freezing studies to date[3,19] confine liquid water at +4C and atmospheric pressure, which equates to a system specific volume $v_s$ of approximately $10^{-4}$ $m^3$/kg. However, these assembly conditions are chosen largely for convenience— the system could also be assembled at higher pressures (lower volumes), forcing the system to follow a different isochore upon cooling. Choice of assembly conditions will vary based on other experimental factors (pressure-sensitivity of a preserved biologic for instance), but at a fundamental level, all assembly conditions (i.e. all isochores) are equally achievable.

In Figure 2, we calculate the phase fraction of ice Ih for four isochores (system volumes) in the temperature range 273 – 190°C. In Figure 2.a, we show the evolution of these phase fractions with temperature as the system traverses three two-phase equilibrium regions, ice Ih-liquid, ice Ih – ice III, and ice Ih – II, and in Figure 2.b we show the corresponding phase fraction – pressure trajectory (note that the intersection of a given curve with the y-axis provides the pressure at which water at 0°C must be confined in order to achieve the isochore experimentally). For further reference, we also include in Figure 2.c the coupling between temperature and pressure in these regions, which the reader may recognize as a portion of the T-P phase diagram, and in Figures 2.d-i we provide the phase fraction on a region-by-region basis for easier reference.

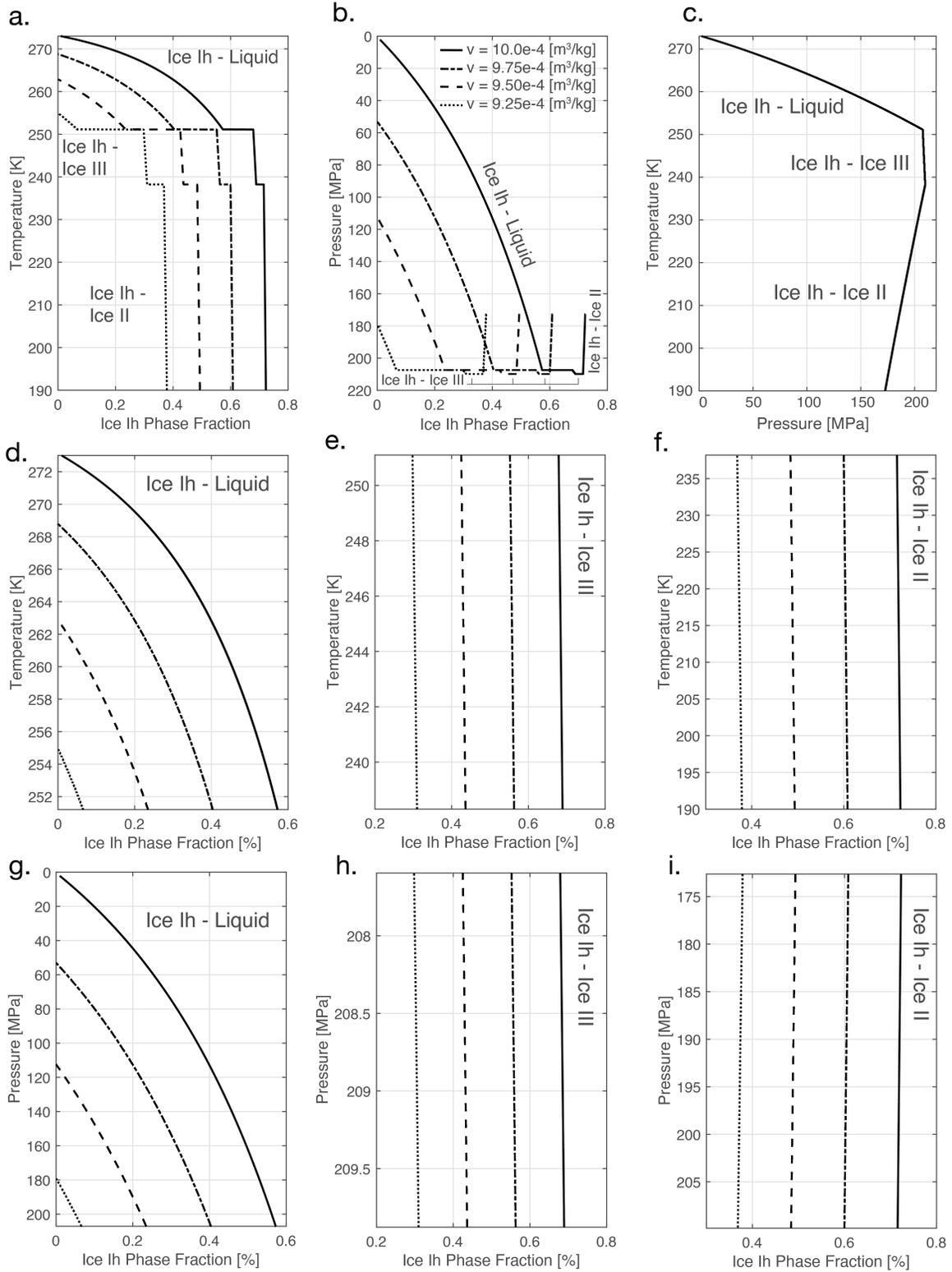

**Figure 2. Evolution of ice Ih phase fraction with temperature for four isochores beginning in the single-phase liquid state.** Isochore volumes for all panels are listed in panel (b) legend. a) Phase fraction – temperature relationship. b) Phase fraction – pressure relationship. c) Temperature-pressure relationship. d-f) Phase fraction – temperature and g-i) phase fraction – pressure relationships for each of the two-phase equilibrium regions traversed in panels a-c).

These phase fraction calculations, which to the author's knowledge have not been performed previously, may be applied in interpretation or design of isochoric cryopreservation protocols intended for cryogenic temperatures, and also to the study of cryovolcanism[20], wherein the significant pressures that emerge due confined aqueous phase change beneath a planetary surface eventually lead to eruption of "molten" aqueous contents.

In assessing the phase fraction evolutions shown in Figure 2, it should be noted further that by assembling an experimental isochoric system at the volumetric eutectic point of water (9.16 x $10^{-4}$ $m^3$/kg), the system may be cooled so as to bypass entirely the ice Ih-liquid two-phase equilibrium region, entering instead directly into the ice Ih – ice III region at low ice-1h phase fractions (less than 30%). For biological matter than can withstand brief exposure to high hydrostatic pressure, this scenario may prove preferable, as the crystallization of denser forms of ice (such as ice III) may prove less damaging than that of ice Ih.

**Further Discussion**

The fundamental similarity between *T-x* and *T-V* phase diagrams (regardless of the number of components in the system) is generalizable to diagrams featuring any arbitrary pairing of an intensive natural variable (such as *T* or *P*) with an extensive variable (such as *x* or *V*); the geometric features are not inherent products of any given physical behavior possessed by multicomponent systems (or isochoric systems for that matter), but instead to the convexity of free energy curves in the free energy – extensive variable space.

This analysis provides two useful conclusions. Firstly, by enforcing any extensive natural variable (e.g. volume, concentration, strain, magnetic moment, polarization, etc.), varied states of heterogeneous multi-phase equilibrium may be achieved in both pure and multi-component substances, analogous to those encountered in classical binary systems, and these varied multiphase equilibria may be mapped by phase diagrams identical in nature and geometry to the common binary *T-x* phase diagrams.

The difficulty of enforcing extensive variables experimentally may vary substantially from application to application, but even restricting our consideration to the simultaneous enforcement of volume and binary concentration, new stable multiphase coexistence regions will emerge in confined binary systems as compared to unconfined binary systems or confined pure systems, and these equilibria will be driven by segregation of component masses between phases *and* segregation of volume between phases. To the author's knowledge, few rigorous *T-V* or *T-V-x* phase diagrams for confined aqueous solutions have been produced to date. Such diagrams may prove invaluable in driving the development of increasingly sophisticated isochoric freezing protocols, and should be prioritized in future work as self-consistent equations of state for common aqueous solutions and the many phases of ice become available.

Secondly, under these same extensive variable conditions, the suite of three-phase invariant reactions typical to binary systems, such as the eutectic transition, may be realized between the condensed phases of pure substances. To the author's knowledge such reactions have not been directly experimentally attempted under isochoric conditions, and exploration of all aspects of such transitions (i.e. nucleation kinetics, crystallization dynamics, etc.) between the myriad phases of pure water could add missing fundamental insight into possible behaviors of the world's most studied substance.

In sum, a new *T-V* phase diagram for water and five of its icy polymorphs has been calculated, providing a reference baseline from which to expand isochoric freezing research into the cryogenic temperature domain. Using this diagram, the fundamental similarity between the phase diagrams (and accordingly phase equilibria) of binary systems with enforced temperature-concentration conditions and pure systems with enforced temperature-volume conditions has been demonstrated, and the resulting "volumetric" eutectic-like reactions are suggested as an intriguing area of continued physical research within the domains of isochoric freezing and water thermodynamics at large. The temperature evolution of the ice Ih phase fraction through three multiphase regions and along four isochores has also been calculated, and may help inform future studies of isochoric freezing, isochoric cryopreservation, and cryovolcanism. The author in parting suggests that the calculation of new phase diagrams under differing enforced natural variables has broad applications outside of isochoric freezing and water thermodynamics, and should be prioritized in any domain wherein phase transitions occur outside of conventional *T-P-x* experimental grounds.

**Methods:**

**Calculation of T-V phase diagram**

*Use of isothermal 2D convex hulls*

In an *n*-dimensional thermodynamic space defined by *n*-1 extensive thermodynamic variable axes plus an internal energy axis, the lower envelope of the convex hull of a cloud of free energy points provides the lowest-free-energy (i.e. globally stable) points. However, when a Legendre transform is performed on the internal energy equation and an intensive variable axis (such as temperature) is introduced, the convex hull can no longer be formed in the *n*-dimensional energy space, because free energy curves become *concave*-down instead of *convex*-down in the intensive variable – free energy axes. Thus, in order to identify stable states in an *n*-dimensional thermodynamic space (here *n* = 3) defined by *n*-1 mixed intensive and extensive variables (here temperature and volume) plus an according thermodynamic potential (here the Helmholtz free energy), (*n*-1)-dimensional convex hulls must be calculated around the extensive variable-free energy (here volume – Helmholtz free energy) curves at each instance of the intensive variable (here temperature).

As such, to calculate the *T-V* phase diagram herein, isothermal convex hulls separated by increments of 0.0654 K were calculated for the volume-Helmholtz free energy curves of all six condensed phases studied here, as represented in Figure 1.a. The points common to the lower convex hulls and the single-phase free energy curves for each phase define the stable points at a given temperature. In order to obtain these points, two steps are required. First, the intersection of the array of points comprising the convex hull and the array of points comprising the single-phase free energy surfaces is computed, giving all common points. Second, the parting line separating the upper envelope of the convex hull from the lower hull is calculated as the line joining the leftmost and rightmost points in the complete hull, and any of the previously computed common points resting above this parting line are discarded. The remaining points represent the stable single-phase regions at a given temperature. Projecting these points onto axes of temperature and volume (i.e. eliminating the free energy axis) yields the colored single phase regions in the phase diagram in Figure 1.c.

*Identification of three-phase invariant lines*

The final step required to complete the diagram is calculation of the three-phase invariants that mark the emergence or disappearance of phases with changing temperature. In the intractable scenario of an infinitely small temperature step, these three-phase configurations would emerge as isothermal convex hulls that have common tangent lines that touch three single-phase free energy curves simultaneously. In the reality of a finite temperature step however, the temperatures at which three-phase invariants occur can be identified by tracking the temperatures at which the number of stable phases changes. When such temperatures are encountered, the bounding volume coordinates of the three-phase invariant lines are obtained by recording the nearest single-phase points to the left and right of the disappearing or emerging phases. As such, the temperature accuracy of the three-phase invariant lines in the Fig. 1c diagram are limited to the accuracy of the temperature step (here chosen to be 0.0654 K).

All calculations and plotting were performed in MATLAB 2021a.

**Acknowledgements:** The author acknowledges no external funding for this work. The author would like to thank the following people: Dr. Baptiste Journaux, for the suggestion that analyses of aqueous isochoric freezing are relevant to cryovolcanism and for assistance in the use of the SeaFreeze framework; Dr. Olivia Dippo, for helpful discussions of early drafts of this manuscript; and to Mr. Anthony Consiglio, for helpful discussions of the concepts at play in this manuscript over the past several years.

**Data Availability:** All data are available from the author upon request.

**Declaration of competing interests:** The author declares to conflicts of interest in the production of this work.